\documentclass[reprint,superscriptaddress, amsmath,amssymb,aps,prapplied,floatfix,longbibliography]{revtex4-1}
\usepackage{textcomp}
\usepackage[english]{babel}

\usepackage{hyperref}
\usepackage{graphicx}
\usepackage{dcolumn}
\usepackage{bm}
\usepackage{hyperref}
\usepackage{graphicx}
\usepackage[dvipsnames]{xcolor}
\usepackage[capitalise]{cleveref}

\hypersetup{pdftoolbar=true,        
    pdfmenubar=true,        
    pdffitwindow=false,     
    pdfstartview={FitH},    
    pdftitle={Kerr Microresonator Soliton Frequency Combs at Cryogenic Temperatures},    
    pdfauthor={Gregory Moille},     
    pdfsubject={CryoComb},   
    pdfcreator={Gregory Moille},   
    pdfproducer={LaTeX}, 
    pdfkeywords={Version 1.0},
    colorlinks=true,       
    linkcolor=blue,          
    citecolor=blue,        
    filecolor=blue,      
    urlcolor=blue}

\begin{document}
\title{Kerr Microresonator Soliton Frequency Combs at Cryogenic Temperatures}
\author{Gregory Moille}
\email{gmoille@umd.edu}
\affiliation{Microsystems and Nanotechnology Division, National Institute of Standards and Technology, Gaithersburg, MD 20899, USA}
\affiliation{Institute for Research in Electronics and Applied Physics and Maryland Nanocenter, University of Maryland,College Park, Maryland 20742, USA}
\author{Xiyuan Lu}
\affiliation{Microsystems and Nanotechnology Division, National Institute of Standards and Technology, Gaithersburg, MD 20899, USA}
\affiliation{Institute for Research in Electronics and Applied Physics and Maryland Nanocenter, University of Maryland,College Park, Maryland 20742, USA}
\author{Ashutosh Rao}
\affiliation{Microsystems and Nanotechnology Division, National Institute of Standards and Technology, Gaithersburg, MD 20899, USA}
\affiliation{Institute for Research in Electronics and Applied Physics and Maryland Nanocenter, University of Maryland,College Park, Maryland 20742, USA}
\author{Qing Li}
\affiliation{Microsystems and Nanotechnology Division, National Institute of Standards and Technology, Gaithersburg, MD 20899, USA}
\affiliation{Institute for Research in Electronics and Applied Physics and Maryland Nanocenter, University of Maryland,College Park, Maryland 20742, USA}
\affiliation{Electrical and Computer Engineering, Carnegie Mellon University, Pittsburgh, PA 15213, USA}
\author{Daron A. Westly}
\affiliation{Microsystems and Nanotechnology Division, National Institute of Standards and Technology, Gaithersburg, MD 20899, USA}
\author{Leonardo Ranzani}
\affiliation{Raytheon BBN Technologies, 10 Moulton Street, Cambridge, MA, 02138, USA}
\author{Scott B. Papp}
\affiliation{Time and Frequency Division, National Institute of Standards and Technology, 385 Broadway, Boulder, CO 80305, USA}
\affiliation{Department  of  Physics,  University  of  Colorado,  Boulder,  Colorado,  80309,  USA}
\author{Mohammad Soltani}
\affiliation{Raytheon BBN Technologies, 10 Moulton Street, Cambridge, MA, 02138, USA}
\author{Kartik Srinivasan}
\email{kartik.srinivasan@nist.gov}
\affiliation{Microsystems and Nanotechnology Division, National Institute of Standards and Technology, Gaithersburg, MD 20899, USA}
\affiliation{Joint Quantum Institute, NIST/University of Maryland, College Park, Maryland 20742, USA}
\date{\today}

\begin{abstract}
    We investigate the accessibility and projected low-noise performance of single soliton Kerr frequency combs in silicon nitride microresonators enabled by operating at cryogenic temperatures as low as 7 K. The resulting two orders of magnitude reduction in the thermo-refractive coefficient relative to room-temperature enables direct access to single bright Kerr soliton states through adiabatic frequency tuning of the pump laser while remaining in thermal equilibrium. Our experimental results, supported by theoretical modeling, show that single solitons are easily accessible at temperatures below 60~K for the microresonator device under study. We further demonstrate that the cryogenic temperature primarily impacts the thermo-refractive coefficient. Other parameters critical to the generation of solitons, such as quality factor, dispersion, and effective nonlinearity, are unaltered. Finally, we discuss the potential improvement in thermo-refractive noise resulting from cryogenic operation. The results of this study open up new directions in advancing chip-scale frequency comb optical clocks and metrology at cryogenic temperatures.
\end{abstract}
\maketitle

\section{Introduction}
Temporal dissipative Kerr soliton pulses generated in nonlinear microresonators are a promising technology for metrological applications of frequency combs~\cite{kippenberg_dissipative_2018}. Their ability to cover a spectral region over an octave~\cite{Li2017,Pfeiffer2017} while operating in a low-noise, phase-stable configuration enables new classes of chip-scale photonics components such as optical frequency synthesizers~\cite{Holzwarth:2000bc8,Spencer:2018cd7}, optical clocks~\cite{Ludlow2015,Drake:2018da4}, and microwave generators~\cite{Fortier:201136d,Lucas:201914b}.  However, reaching soliton states in practice can be challenging. These states lie on the red-detuned side of the microresonator's pump resonance mode~\cite{Herr2013,Godey2014}, for which thermo-refractive effects limit stability~\cite{carmon_dynamical_2004}. This can make accessing soliton states difficult in thermal equilibrium, depending on the properties of the system~\cite{Li2017}. Larger scale resonators, such as MgF\textsubscript{2} crystalline devices~\cite{Herr2013} and  SiO\textsubscript{2} microresonators~\cite{Yi2015}, have sufficiently large thermal conductivity and sufficiently slow thermal time constants to enable soliton generation through slow adjustment of the pump laser frequency to the appropriate detuning level. In chip-integrated, planar microresonators, thermal timescales are faster and other approaches are typically required.  Fast frequency shifting via single-sideband modulators~\cite{Briles:20188c6}, integrated micro-heaters with fast response times~\cite{joshi_thermally_2016}, abrupt changes to the pump power level~\cite{Brasch2016a}, phase modulation of the pump~\cite{cole_kerr-microresonator_2018}, and an auxiliary laser for providing temperature compensation~\cite{zhang_sub-milliwatt-level_2019} have all been used, while resonators with sufficiently low optical absorption can obviate the need for such methods~\cite{liu_photonic_2019}.

\begin{figure}[t]
  \includegraphics{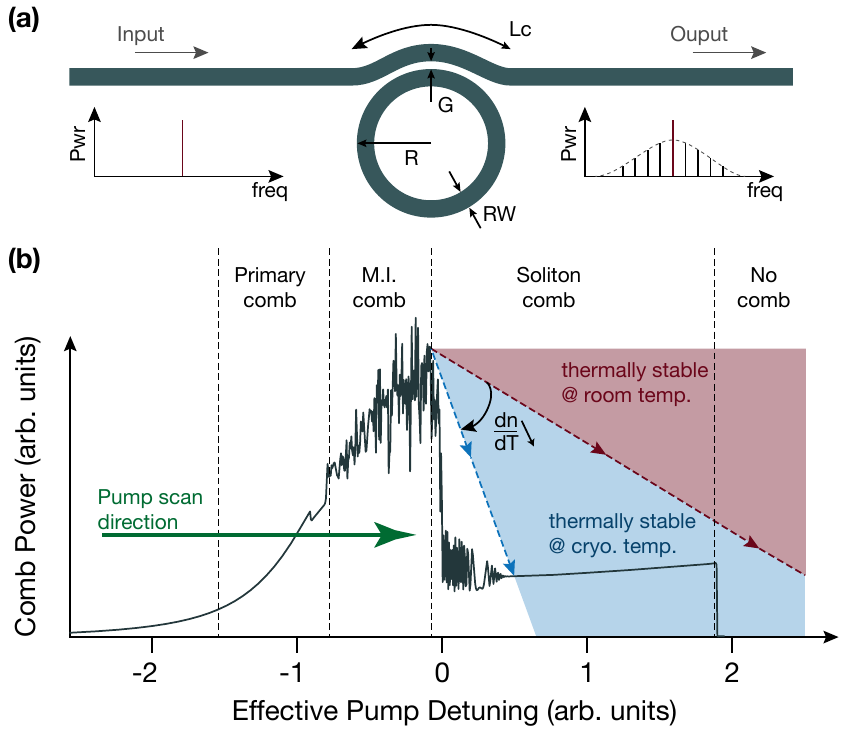}
  \caption{\label{fig:Schematic} (a) Schematic of the microring studied, where a single frequency, continuous-wave laser at the input results in a soliton microcomb at the output thanks to the $\chi^{(3)}$ nonlinearity of the resonator. RW: ring width, G = gap, R: radius, L\textsubscript{c}: coupling length (b) Thermal accessibility of soliton states. The black line shows the LLE-simulated comb power (i.e. the integrated power in the comb with the pump line filtered-out) as a function of laser-pump mode detuning, while the red and blue dashed lines represent a range of solution to the thermo-refractive model as temperature (hence $\frac{\partial n}{\partial T}$) is decreased. The intersection of the thermal model solution and the soliton step of the LLE simulation determines whether that soliton state is thermally stable. MI: modulation instability.}
\end{figure}


%

Here, we consider another approach, so far unexplored, which is to strongly reduce the thermo-refractive coefficient $\frac{\partial n}{\partial T}$ of the resonator, making soliton states accessible with slow frequency tuning of the pump laser (\cref{fig:Schematic}). This is accomplished by operating silicon nitride (Si$_3$N$_4$) microresonators at cryogenic temperatures ($T\leq60$~K), where $\frac{\partial n}{\partial T}$ drops significantly~\cite{Elshaari:2016f7b} (e.g. two orders of magnitude reduction at 10~K relative to room temperature). The thermal expansion coefficient for dielectric thin films also goes down with temperature~\cite{Ekin}, though for silicon nitride, it starts $\approx$~10$\times$ lower than $\frac{\partial n}{\partial T}$~\cite{kaushik_wafer-level_2005} and is henceforth not considered.  After experimentally confirming this large decrease in $\frac{\partial n}{\partial T}$, we examine resonator dispersion, quality factor, and parametric oscillation threshold, and find that they are largely unaffected by the change in temperature.  We then confirm that a reduced thoermo-optics coefficient enables straightforward access to single soliton states, and examine how this accessibility changes as temperature, and thus $\frac{\partial n}{\partial T}$, is increased. Our experimental results are in good correspondence with the theory developed in Ref.~\cite{Li2017}, where the Lugiato-Lefever equation (LLE) for modeling Kerr dynamics is combined with a simple thermal model. Outside of this first demonstration of soliton cryo-combs, we theoretically consider the implications of cryogenic operation on soliton stability. In particular, we consider how reduced thermorefractive effects can lower carrier-envelope offset frequency noise~\cite{Drake:2019af1} by several orders of magnitude.
\section{Results and Discussion}

\subsection{Thermal Accessibility of Soliton States: Model}

To understand the thermal accessibility of Kerr soliton states in a microresonator, schematically described in \cref{fig:Schematic}(a), we employ the theoretical model presented in Ref.~\cite{Li2017} as a base for our study. Due to the large difference in timescales for Kerr and thermal dynamics, thermo-refractive effects can be added to the LLE model in a two-step fashion.  First, the LLE is simulated to determine the behavior of the comb power as a function of pump laser detuning with respect to its adjacent cavity mode, such that a positive detuning corresponds to a red-detuned pump laser (see \cref{fig:Schematic}(b)). As the pump frequency is swept across the transition from chaotic, modulation instability (MI) states to the soliton regime, the large drop in comb power results in a sudden temperature change, with a resulting shift of the cavity resonance due to the thermo-refractive effect. A simple thermal model results in a linear solution with a slope $K_{\rm eff}\propto (\frac{\partial n}{\partial T})^{-1}$. When the solution of the thermal model intersects a soliton step, the shift of the cavity mode due to the temperature change upon entering the soliton regime is small enough that the system remains in a soliton state.  Since $\frac{\partial n}{\partial T}$ is temperature-dependent, soliton steps which are inaccessible at room temperature can be accessed at cryogenic temperatures as shown in \cref{fig:Schematic}(b) for which $\frac{\partial n}{\partial T}$ is sufficiently small.

\begin{figure}[t]
  \includegraphics{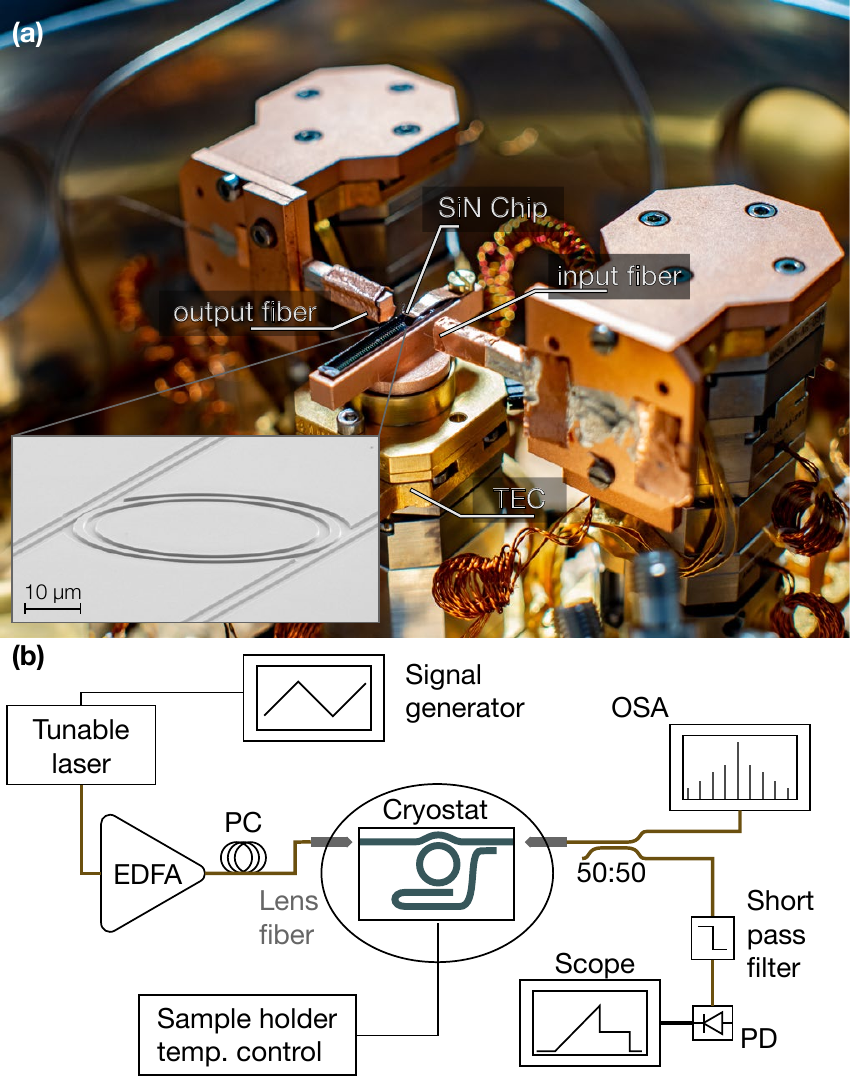}
  \caption{\label{fig:CryostatPic}(a) Photograph of the cryostat interior with input/output fiber coupling to the Si\textsubscript{3}N\textsubscript{4} chip.The coupling to the ring is made with two waveguides to extract different portions of the comb spectrum, and are combined before the output port through an on-chip dichroic element.  The chip sits on a copper mount affixed to a feedback-controlled heater stage that enables the temperature to be varied from 6~K to 90~K without adjusting the cooling power to the cryostat.  The inset is a SEM image of an individual microring resonator. (b) Experimental setup for soliton generation at cryogenic temperature. OSA: Optical Spectrum Analyser, PD: Photo-Diode, PC: Polarization Controller, EDFA: Erbium-Doped Fiber Amplifier}
\end{figure}

\subsection{Thermorefractive Coefficient, Cavity $Q$, and Comb Threshold vs. Temperature}
The device we investigate is a microring resonator (inset of \cref{fig:CryostatPic}(a)) made of 617 nm thick Si\textsubscript{3}N\textsubscript{4} with a 3~{\textmu}m SiO\textsubscript{2} bottom layer above a Si substrate, without any top cladding layer. The other geometrical dimensions of the microring are an outer ring radius $R$ = 23 {\textmu}m and ring width $RW$ = 1850 nm, resulting in a dispersion profile that supports octave-spanning soliton comb states, as accessed through fast frequency sweeping in ref.~\cite{Briles:20188c6}. Such rapid pump frequency adjustment was a necessity due to the room-temperature thermal dynamics preventing single soliton accessibility through adiabatic tuning of the frequency. We placed the chip inside a 20~cm diameter chamber within a closed-cycle cryostat with a base temperature $T\leqslant 5$~K (\cref{fig:CryostatPic}). The sample is mounted on a copper sample holder that is affixed to a temperature-controlled stage which incorporates a resistive heater with an embedded temperature sensor and closed-cycle control. Hence, the temperature of the sample can be tuned between 6~K and 100~K  without adjusting the cooling power to the cryostat. The sample is optically addressed using lensed optical fibers, and special care was taken to thermally sink both the fibers and the sample, to limit adverse heating effects when significant optical powers are coupled into the system.
%
\begin{figure}[t]
  \includegraphics{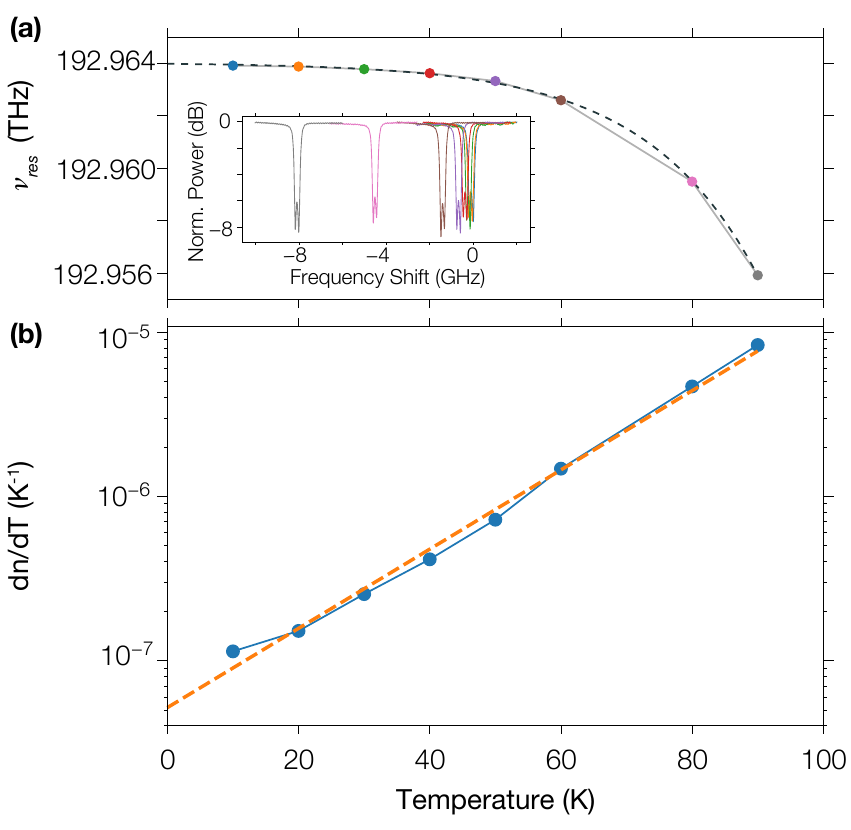}
  \caption{\label{fig:dndT} (a) Spectral shift of the microring resonator pump mode with  temperature. The dashed line corresponds to the exponential fit used to retrieve $\partial n/\partial T$. The inset represents the measured resonance profile, a doublet that is fit using the model from Ref.~\cite{borselli_beyond_2005}. (b) Variation of the thermo-refractive coefficient of the system with temperature. The dashed line corresponds to the exponential trend. Uncertainties are within the size of the markers.
  }
\end{figure}

\begin{figure*}[t]
  \includegraphics{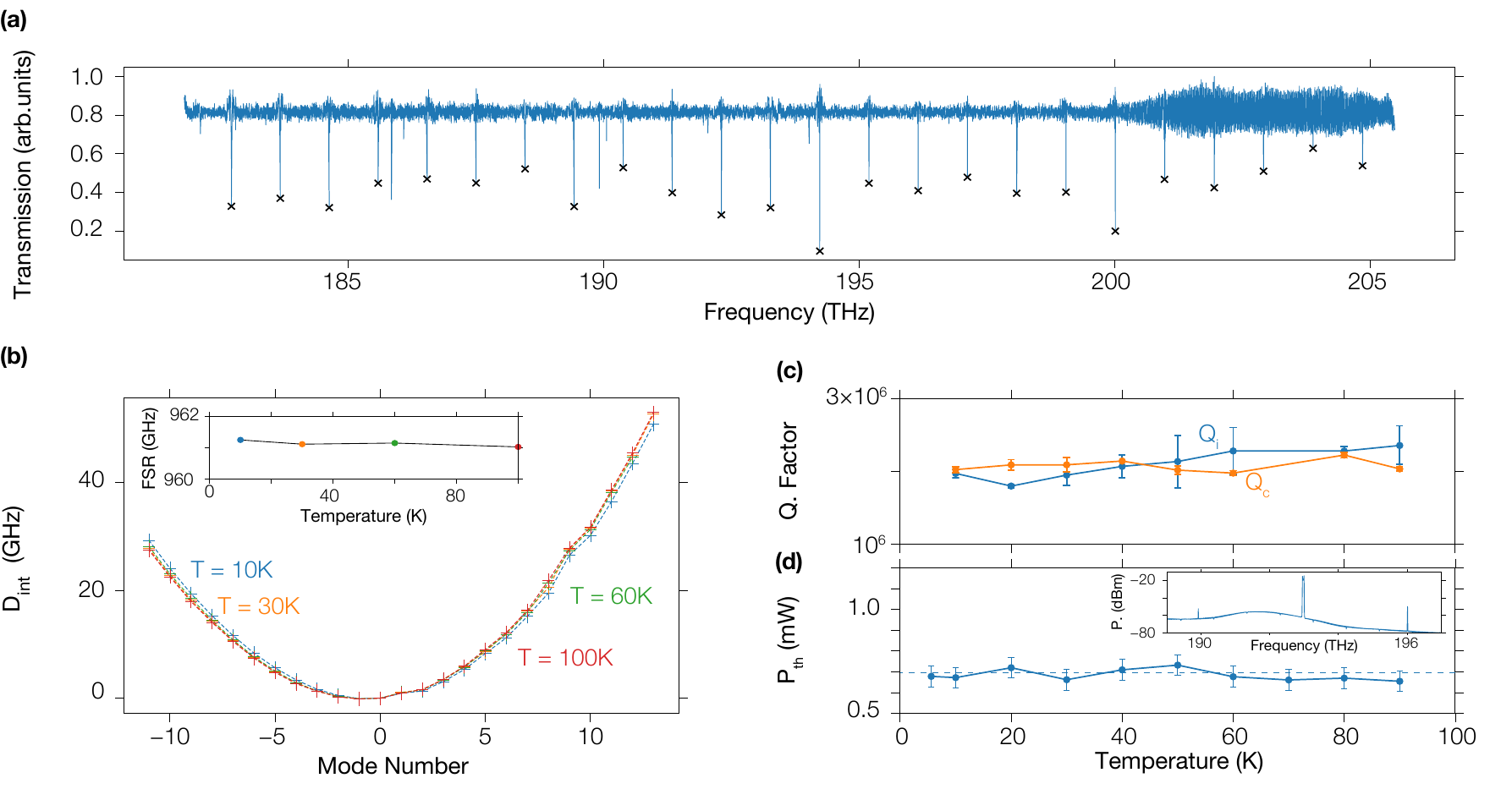}
  \caption{\label{fig:QPthDisp} (a) Linear transmission of the ring resonator at 60~K. The crosses mark the first order mode family of interest for broad soliton comb generation. (b) Integrated dispersion obtained from the linear transmission data such as that in (a) as a function of mode number and at different temperatures. For the frequency range considered, the dispersion is nearly quadratic. The inset corresponds to the change in the extracted FSR with temperature. Uncertainties are within the size of the markers.  (c) Behavior of the intrinsic (blue) and coupling (orange) quality factor with temperature. Error bars represent 95~$\%$ confidence intervals from a nonlinear least square fit to the data. (d) Threshold power $P_{th}$ for optical parametric oscillation (OPO) as a function of temperature, where the error bars are one standard deviation values due to variation in the fiber insertion loss. The inset corresponds to the optical spectrum of the first OPO side-bands obtained at $P_{th}$.}
\end{figure*}

We first experimentally examine the thermo-refractive coefficient, which is retrieved by measuring the frequency shift of a resonance $\frac{\partial \nu}{\partial T}$ with temperature from 10~K up to 90~K (\cref{fig:dndT}(a)), in particular probing the mode used later as the pumped one to create a microcomb. From the mode frequency shift, and its relation with the thermo-refractive coefficient
$\frac{\partial n}{\partial T} \equiv \left( \frac{\partial \nu_{0}}{\partial \tilde{n}}\right)^{-1}\frac{\partial \nu_{0}}{\partial T}$, with $\frac{\partial\nu}{\partial n}=-92.57$~THz (calculated with a Finite-Element Method (FEM) eigensolver), $\frac{\partial n}{\partial T}$ of the system is determined.  The benefit of going to cryogenic temperature is clear as the thermo-refractive coefficient of the system drops by 2 orders of magnitude in going from 90~K to 10~K (\cref{fig:dndT}(b)), and exhibits more than a factor two difference between 90~K and its value reported at room temperature $\left.\frac{\partial n}{\partial T}\right|_{300 \rm{K}} = 2.45 \times 10^{-5}$~K\textsuperscript{-1}~\cite{arbabi2013}. These results are consistent with data reported on the cryogenic behavior of the thermo-refractive coefficient for both Si\textsubscript{3}N\textsubscript{4} and SiO\textsubscript{2}~\cite{Elshaari:2016f7b}.

To check that the primary influence of cryogenic temperature on soliton accessibility is through the reduced $\frac{\partial n}{\partial T}$, it is essential to verify that no other parameters critical to soliton formation and stability significantly vary with temperature. First, we consider whether temperature influences the resonator dispersion, through measurement of its mode frequencies from $\approx$~180~THz to $\approx$~205~THz (\cref{fig:QPthDisp}(a)). These measurements allow us to retrieve the free spectral range (FSR) for the targeted mode family, as well as the integrated dispersion, given by $D_{\rm int} = \nu_\mu - (\nu_{0} + \rm{FSR}\times$~$\mu )$ where $\mu$ represents the mode number relative to the pumped mode frequency $\nu_{0}$, and $\nu_\mu$ is the frequency of the $\mu$\textsuperscript{th} mode. For both, no significant change is observed with temperature (\cref{fig:QPthDisp}(b)), which indicates that the thermo-refractive coefficient is quasi non-dispersive in the frequency band of interest. Given the sensitivity of dispersion on the resonator geometry, this suggests that the ring dimensions are also largely unchanged with temperature, and supports our assumption that thermal expansion effects can be neglected over the temperature range studied. This claim is further supported by measurement of the intrinsic and coupled quality factor, which are found to be unaltered from 10~K to 90~K (\cref{fig:QPthDisp}(c)), suggesting that the geometry is not changing significantly as the resonator-waveguide coupling is sensitive to relatively small geometric changes. Finally, by increasing the pump laser power while keeping a fixed detuning on the blue side of the pump cavity resonance, we observe optical parametric oscillation (OPO), as seen in the inset to \cref{fig:QPthDisp}(d).  We find that the threshold power for OPO is approximately constant with temperature. As the quality factor and the power-threshold are unchanged with temperature, we can conclude that non-linear coefficient $n_2$ of the Si\textsubscript{3}N\textsubscript{4} remains unchanged; hence only the thermo-refractive coefficient is impacted by the cryogenic temperatures.

\begin{figure*}[t]
  \includegraphics{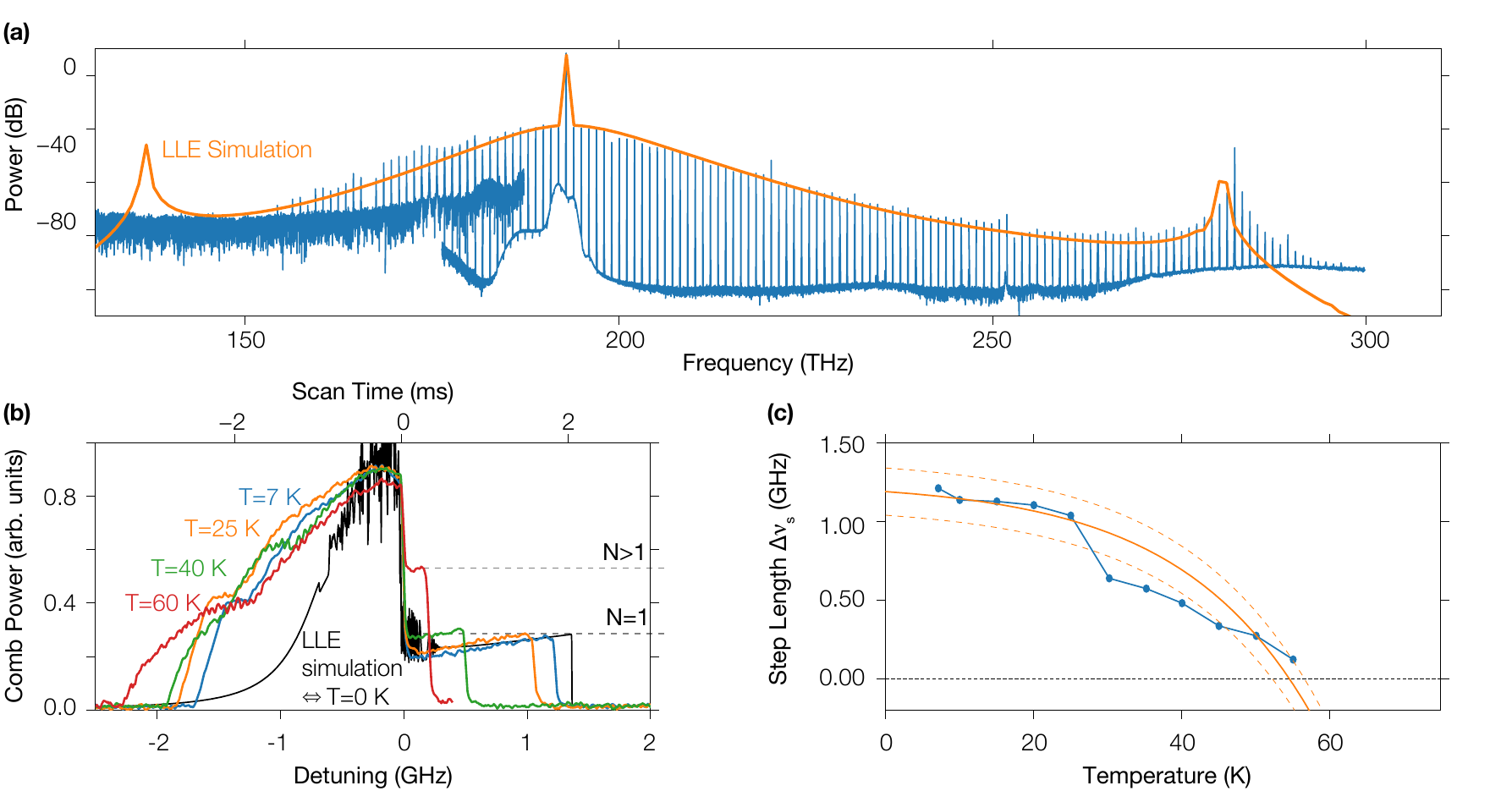}
  \caption{\label{fig:SolitonStep} (a) Measured (blue) and simulated (orange) single soliton frequency comb obtained at $T\approx$~7~K with a pump power of 100 mW in the waveguide. The spectrum is acquired using two optical spectrum analyzers, with the noise floor of the instrument used for acquiring the lower frequency section about 20 dB higher. (b) Comb power versus detuning $\nu_0 - \nu_{\rm pmp}$ obtained at different temperatures. The black line corresponds to the LLE simulation. (c) Measured soliton step length (blue) versus temperature, compared to theoretical values assuming nominal parameters from the experiment (solid orange line). In addition, we plot a range of theoretical values (dashed orange lines) defined by $\pm$1~dB variation in the pump power, which represents the observed fluctuation in insertion loss during the experiment.}
\end{figure*}

\subsection{Soliton Generation and Step Length vs. Temperature}

We measured the soliton dynamics of the system by pumping the resonance at $\nu_{0} \approx 192.96$~THz that was used to characterize the quality factor, the power-threshold, and the thermal shift.  Applying an in-waveguide power of $P_{\rm pmp}\approx~100$~mW, a single soliton state is easily reached through adiabatic tuning of the pump laser frequency due to the drop of the thermo-refractive coefficient, and the spectrum of the resulting microcomb is reported in \cref{fig:SolitonStep}(a), exhibiting the characteristic and expected $sech^2$ envelope. The measured microcomb spectrum is in good agreement with the spectral envelope predicted by simulations of the LLE performed using an open-source software package~\cite{Moille:2019291}. One exception is the lack of a low frequency dispersive wave around 135~THz, which is not present experimentally. We believe that this is most likely a technical issue that is a result of our measurement. Due to the small space available in between the cryostat radiation shield and fiber feedthrough that connects the vacuum space to the outside laboratory, the SMF28e fibers~\cite{NIST_disclaimer} running to and from the comb chip undergo tight bends. As has been seen in other literature~\cite{Kuo:2018f66}, bend diameters below 40~mm can cut-off long wavelength light below 2~{\textmu}m (150~THz).

We examined a range of cryogenic temperatures for which a single soliton was accessible and found the soliton step width for the corresponding temperature. For this experiment, we measured the comb power, \textit{i.e.}, the output power when the pump is filtered out, as a function of temperature and for an in-waveguide pump power of $P_{\rm pmp}\approx40$~mW (\cref{fig:SolitonStep}(b)). %
This power level was chosen based upon simulations that indicate a favorable operating point for single soliton generation.  The comb power traces (\cref{fig:SolitonStep}(b)) indicate a transition from states accessible on the blue-detuned side of the resonance (comb power increasing) to the abrupt soliton steps on the red-detuned side (we note that the detector bandwidth of 500 kHz precludes observation of the fast intensity variations expected for chaotic states on the blue-detuned side, as seen in the LLE simulation). It is clear from these traces that the single soliton step is observable up to 60 K, and the step width is wider for lower temperatures, making the soliton regime more accessible.%
It is clear from the comb power traces in \cref{fig:SolitonStep}(b) that the single soliton step is observable up to 60~K, and the step width is wider for lower temperatures, making the soliton regime more accessible.  At a temperature of 60~K, a soliton step at higher comb power is observed, corresponding to a higher number of soliton pulses in the cavity. We emphasize the time scale of the pump frequency sweep (milliseconds), which is orders of magnitude slower than the thermal lifetime in our system, usually on the order of tenths of a microsecond~\cite{Brasch2016a}. In ref.~\cite{Li2017}, it was shown that under such slow frequency sweeps, only multi-soliton states were accessible at room-temperature unless a very specific condition of temperature compensation via coupling to an adjacent optical mode was possible. Here, we consider soliton accessibility at cryogenic temperature.

To further investigate the single soliton dynamics at cryogenic temperatures, we compare the measured single soliton step length with temperature and the theoretical one obtained through the thermal stability study presented in ref~\cite{Li2017}. Based on that work, one can write the soliton step length as:

\begin{equation}
    \Delta\nu_{\rm s} (T) = \Delta\nu_{\rm s}^0 - (1-\eta)K_{\rm eff}^{-1}(T)
\end{equation}

\noindent with $\Delta\nu_{\rm s} (T)$ the soliton step length at the temperature $T$ and $\Delta\nu_{\rm s}^0$ is the theoretically predicted step length from the LLE (in absence of thermal effects, i.e., at $T=0$), $\eta$ the ratio between the peak of the modulation instability (MI) comb power and the soliton step power, and $K_{\rm eff}$ describes the thermal frequency shift such that $K_{\rm eff}^{-1} = 2\frac{\omega_{0}t_{\rm R}}{n_{\rm g} K_{\rm c}}\frac{\kappa_a}{\kappa}\times\frac{\partial n}{\partial T}$, where $\kappa_a/\kappa$ is the ratio of the linear absorption rate to the total loss rate. Other parameters include a group index $n_{\rm g} = 2.0669$, round-trip time $t_{\rm R}=0.996$~ps  obtained from FEM simulation, $\Delta\nu_{\rm s}^0=1.23$~GHz obtained from LLE simulation, and thermal conductance $K_{\rm c} =2.86\times10^{-4}$~W/K. To match our experimental data, the prefactor in front of $\frac{\partial n}{\partial T}$ in $K_{\rm eff}^{-1}$ needs to be about a factor of two larger than that calculated in Ref.~\cite{Li2017}. This difference is perhaps not surprising, because of the different thermal environment (microring is now in vacuum instead of air), the potential change in the thermal conductivity of  Si\textsubscript{3}N\textsubscript{4} and SiO\textsubscript{2} at cryogenic temperatures, and varying fraction of loss due to absorption from device to device. From the thermal/LLE model, we retrieve the theoretical behavior of the soliton step length (\cref{fig:SolitonStep}(c)) which is consistent with the experimental results (particularly when accounting for potential variation in the pump power due to changes in coupling; see dashed lines in \cref{fig:SolitonStep}(c)). It also shows that above 60~K, the single soliton state is not thermally stable anymore ($\Delta\nu_{\rm s}<0$), which also is consistent with the experimental observation.

\subsection{Potential Impact of Reduced Thermorefractive Noise on Soliton Microcomb Stability}

Finally, an essential consequence of the drop in the thermo-refractive coefficient at cryogenic temperatures is the potential to reduce thermo-refractive noise. Indeed, the resonance frequencies of the different modes of the microring depend on the material refractive index. Hence, thermal fluctuation leads to frequency noise through the thermo-refractive coefficient. The mean thermal fluctuation can be derived from thermodynamics as $\langle \delta T^2 \rangle = \frac{k_{\rm B} T^2}{C V_{\rm m} \rho}$  with $C$ the heat capacity, $V_{\rm m}$ the optical mode volume of the mode $m$, $k_{\text{B}}$ is Boltzmann's constant, and $\rho$ the density of the resonator material. The thermal fluctuations are directly linked to the variation of the resonance frequency $\frac{\delta \nu}{\nu} = -n^{-1}\frac{\partial n}{\partial T}\delta T$, which can be reformulated as the spectral density of optical fluctuations:
\begin{align}
\label{eq:Su}
S_{\nu} = \left(\nu_0 \cdot n^{-1}\frac{\partial n}{\partial T}\right)^2 S_{\rm T}
\end{align}

\noindent with $S_{\rm T}$ the spectral density of thermal fluctuations. This last physical quantity has been the subject of investigation in different works, and several analytical models have been reported. The two models that will be used in this work, and have been demonstrated to match Si\textsubscript{3}N\textsubscript{4} experimental data well for different Fourier frequency spectral windows~\cite{Huang:20198ab} are the approximation of a homogeneous cavity within a heat bath~\cite{Kondratiev2018}, and the approximation of the thermal decomposition method~\cite{matsko2007}. The former can be written as:
\begin{align}
\label{eq:ST1}
    S_{\rm T}(\omega) =&\frac{k_B T^2}{\sqrt{\pi^3 \kappa \rho C \omega}}\sqrt{\frac{1}{2m}}\times\nonumber\\
    &\frac{1}{R\sqrt{d_r^2-d_z^2}}\frac{1}{\left(1 + \left(\omega \tau_d\right)^{3/4}\right )^2}
\end{align}
\noindent with $R=23$~{\textmu}m the ring radius,  $d_z$ and $d_r$ the half-width of the fundamental mode with the azimuthal mode order $m=159$ in $r$ and $z$ directions respectively, $\tau_d = \frac{\pi ^{3/4}}{4^{1/3}}\frac{\rho C}{\kappa}d_z^2$, $\omega$ the Fourier frequency and $\kappa$ the thermal conductivity of the material. We use  thermal properties one can find in the literature, including specific heat capacity $C=800$~J$\cdot$kg\textsuperscript{-1}$\cdot$K\textsuperscript{-1}, $\kappa=30$~W$\cdot$m\textsuperscript{-1}$\cdot$K\textsuperscript{-1}, and density $\rho=3.29\times 10^3$~kg$\cdot$m\textsuperscript{-3}.

The second model used here can be written as:
\begin{align}
\label{eq:ST2}
S_{\rm T}(\omega) = &\frac{k_B T^2 R^2}{12\kappa V_{\rm m}}\times\nonumber\\&\left( 1+ \left(\frac{R^2 \rho C \omega}{3^{5/2}\kappa}\right)^{3/2}  + \frac{1}{6}\left( \frac{R^2\rho C \omega}{8 m^{1/3}\kappa}\right)^2\right)^{-1}
\end{align}

\noindent with $V_{\rm m}=1\times10^{-16}$~m\textsuperscript{-3} the effective volume.

\begin{figure}[t]
  \includegraphics{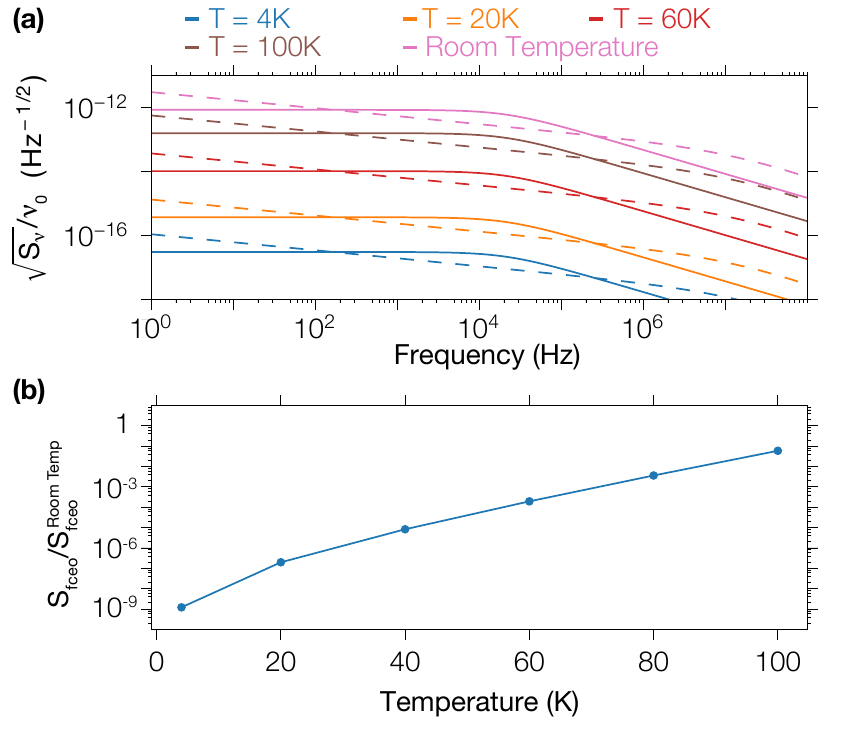}
  \caption{\label{fig:Noise} Calculation of (a) the fractional spectral density of optical frequency fluctuations $\sqrt{S_{\nu}}/\nu_0$ and (b) ratio of the spectral density of soliton comb carrier-envelope offset frequency fluctuations $S_{\rm fceo}$ at a given temperature to the value at room temperature. We assume the experimental values of $\partial \nu_0/\partial T$ from \cref{fig:dndT} and other parameters from Ref.~\cite{Drake:2019af1}.}
\end{figure}

From \cref{eq:Su,eq:ST1,eq:ST2}, the improvement at cryogenic temperatures is obvious, as the spectral density of frequency fluctuation scales with $\left(\frac{\partial n}{\partial T}\right)^2 T^2$, and \cref{fig:Noise}(a) shows an improvement of four orders of magnitude in the normalized spectral density of optical fluctuation $\sqrt{S_\nu}/\nu_0$ compared to its value at room temperature. However, even if the spectral density of noise is largely reduced by working at cryogenic temperature, the $Q$ of Si\textsubscript{3}N\textsubscript{4} rings such as those we study is generally too low to consider them as viable candidates for reference cavities~\cite{deHond:17}. Going forward, recent improvements in the $Q$ of thick  Si\textsubscript{3}N\textsubscript{4}~\cite{xuan2016,ji_ultra-low-loss_2017,liu_ultralow-power_2018} resonators, as well as earlier work on thin Si$_3$N$_4$ resonators with quality factors approaching 10$^8$~\cite{spencer_integrated_2014}, suggest future potential to combine such higher $Q$ with cryogenic operation towards reference cavity applications.

Applications of octave-spanning soliton frequency combs typically require a stabilized carrier-envelope offset frequency ($f_{\rm ceo}$), which is often determined through the $f$-2$f$ beat-note detection scheme. However, fluctuations in this beat-note can arise from thermal noise, setting a limit on the metrological applications of such combs~\cite{Drake:2019af1}. From Ref.~\cite{Drake:2019af1}, one can link the spectral density of thermal noise to that of $f_{\rm ceo}$, such that:
\begin{equation}
    S_{\rm fceo}(\omega) = p^2 \eta_{\rm rep}^2 S_{\rm T}(\omega)
\end{equation}

\noindent with $p=192$ the pumped comb mode number and $\eta_{rep}$ given by:
\begin{equation}
    \eta_{\rm rep} = \left(\frac{\partial f_{\rm rep}}{\partial\nu_{0}} - \frac{\partial f_{\rm rep}}{\partial \Delta}\right)\frac{\partial \nu_0}{\partial T}
\end{equation}

Like the spectral density of optical frequency fluctuations, the spectral density of $f_{\rm ceo}$ fluctuations scales with  $\left(\frac{\partial n}{\partial T}\right)^2 T^2$ and should thus strongly benefit from cryogenic temperature operation, both from the reduction in temperature and the drop of the thermo-refractive coefficient. This leads to a predicted improvement in the density of noise fluctuations of close to nine orders of magnitude (\cref{fig:Noise}(b)).

\section{Conclusion}

In conclusion, we have demonstrated soliton generation at cryogenic temperature in Si\textsubscript{3}N\textsubscript{4} microrings. This is enabled by the drop in the thermo-refractive coefficient, which we have measured to be more than two orders of magnitude smaller compared to the room temperature value, quenching the thermal bistability and leading to thermally stable soliton states.  In addition, measurement of critical parameters such as quality factors, dispersion, and threshold power have been performed at different temperatures, and lead to the conclusion that the cryogenic temperature primarily modifies the thermo-refractive coefficient. Moreover, the ability to strongly suppress thermorefractive effects may enable experimental verificatiosn of a number of theoretical predictions that are based on studies of the Kerr dynamics alone. We have further validated the theory introduced in Ref.~\cite{Li2017} by varying the temperature of the device, which alters the thermo-refractive coefficient and results in a change of the soliton step length consistent with the theory. Finally, we have theoretically investigated the expected reduction in noise, both for the spectral density of optical frequency fluctuations, important for reference cavity applications, and the spectral density of carrier-envelope offset frequency fluctuations, which has a significant impact on the stability of phase-locked octave-spanning microcombs. For both figures of merit, we expect an improvement of several orders of magnitude, thanks to both the drop in temperature and the drop in thermo-refractive coefficient. This suggests that other noise sources may be the ultimate limiting noise factor when Si$_3$N$_4$ microrings are operated at cryogenic temperature. Though cryogenics adds a level of complexity beyond experiments done in ambient, the potential for such temperatures to be naturally reached in some satellite applications, as well as recent advances in miniaturized cryocoolers~\cite{conrad2017}, indicates significant potential impact in the implementation of low-noise and stable compact clocks and metrology system.

\begin{acknowledgments}

 This work is supported by the DARPA DODOS, ACES, and NIST-on-a-chip programs. G.M., X.L., and Q.L. acknowledge support under the Cooperative Research Agreement between the University of Maryland and NIST-PML, Award no. 70NANB10H193. The authors thank Tara Drake from NIST Boulder and Amit Agrawal from NIST Gaithersburg for helpful discussions.
\end{acknowledgments}
 \bibliography{2019CryoComb}
 \end{document}